\documentclass{rspublic}

\begin{document}

\title{Resummation and the semiclassical theory of spectral statistics}

\author[J. P. Keating and S. M\"uller]{Jonathan P. Keating$^1$ and Sebastian M\"uller$^2$}

\affiliation{$^1$School of Mathematics, University of Bristol, Bristol BS8 1TW, United Kingdom,
$^2$Cavendish Laboratory, University of Cambridge, J J Thomson
Avenue, Cambridge CB3 0HE, United Kingdom}

\label{firstpage}

\maketitle


\begin{abstract}{quantum chaos, spectral statistics, semiclassical approximation, periodic orbit theory, random matrix theory}

We address the question as to why, in the semiclassical limit,
classically chaotic systems generically exhibit universal quantum
spectral statistics coincident with those of Random Matrix Theory.
To do so, we use a semiclassical resummation formalism that
explicitly preserves the unitarity of the quantum time evolution
by incorporating duality relations between short and long
classical orbits. This allows us to obtain both the
non-oscillatory and the oscillatory contributions to spectral
correlation functions within a unified framework, thus overcoming
a significant problem in previous approaches. In addition, our
results extend beyond the universal regime to describe the
system-specific approach to the semiclassical limit.

\end{abstract}

\maketitle

\section{Introduction}

One of the central ideas in the field of Quantum Chaos is that the
quantum energy spectra of classically chaotic systems exhibit
universal fluctuation statistics (Bohigas {\it et al.} 1984,
McDonald \& Kaufman 1979, Casati {\it et al.} 1980, Berry 1987):
once their mean energy level densities are scaled to be equal, the
spectral statistics of different systems coincide, forming a small
number of symmetry-dependent universality classes. This idea is
supported by overwhelming experimental and numerical evidence for
systems as diverse as atomic nuclei, Rydberg atoms and quantum
billiards (St\"ockmann 1999). Theoretically, it emerges in the
semiclassical limit, $\hbar \rightarrow 0$, when spectral
correlations are measured on the scale of the mean energy level
separation.

The fundamental reasons for universality are still not fully
understood. Nevertheless, a phenomenological description of the
specific universal forms taken by spectral statistics has been
developed based on the following insight: accepting universality,
one may consider ensemble averages rather than individual chaotic
systems. This idea lies at the heart of random-matrix theory
(RMT), developed by Wigner and Dyson to describe spectral
statistics in atomic nuclei (Wigner 1959, Haake 2001). In RMT, one
averages over ensembles of all matrix representations of
Hamiltonians belonging to the same symmetry class, with, for
convenience, a Gaussian weight. The pertinent ensemble for systems
without any symmetries, comprising of all complex Hermitian
matrices, is referred to as the Gaussian Unitary Ensemble (GUE);
for dynamics whose sole symmetry is time-reversal invariance the
Gaussian Orthogonal Ensemble (GOE) of real-symmetric matrices is
the appropriate one. Ensemble averages then yield predictions for
measures of spectral statistics such as the two-point correlation
function of the level density of energy levels $E_n$,
$\rho(E)=\sum_n \delta(E-E_n)$:
\begin{equation}
R(\epsilon)\equiv
\frac{1}{\overline{\rho}^2}\left\langle\rho\left(E+\frac{\epsilon}{2\pi\overline{\rho}}\right)
\rho\left(E-\frac{\epsilon}{2\pi\overline{\rho}}\right)\right\rangle-1.
\end{equation}
Here $\langle.\rangle$ denotes an ensemble average, replacing an
average over the centre energy $E$ for individual systems, and
$\overline\rho$ is the mean level density. In the large-matrix
limit, the random-matrix predictions involve power series of
non-oscillatory and oscillatory contributions with coefficients
$c_n,d_n$ (Heusler {\it et al.} 2007) depending on the symmetry
class:
\begin{equation}
\label{Rseries} R(\epsilon) ={\rm
Re}\sum_{n=2}^\infty\left(c_n+d_n{\rm
e}^{2i\epsilon}\right)\left(\frac{1}{\epsilon}\right)^n\;.
\end{equation}
For the GUE, $c_2=-d_2=-1/2$ and $c_n=d_n=0$ for all $n>2$.

Establishing precisely when and why random matrix theory describes
the spectral statistics of individual classically chaotic systems
is one of the central problems in the field.  To illustrate its
difficulty, we remark that several fully chaotic systems are
known, such as the cat maps (Keating 1991) and geodesic motion on
compact arithmetic surfaces of constant negative curvature
(Bogomolny {\it et al.} 1992), for which the spectral statistics
do not coincide with any of the random-matrix forms; and there are
examples of perturbed cat maps where the spectral statistics do
coincide with one of the random-matrix forms, but not the form one
would predict based on the symmetries of the classical dynamics
(Keating \& Mezzadri 2000).  In all of these cases the origin of
the unexpected behaviour lies in the existence of arithmetical
quantum symmetries that have no classical counterpart.  There are
currently no rigorously established necessary or sufficient
conditions for the quantum spectrum of any individual classically
chaotic system to exhibit random matrix statistics.  For this
reason, the relationship is usually conjectured to hold for {\it
generic} systems, where the term `generic' is sufficient to cover
experimental applications, but as yet has not been given a precise
mathematical definition.

The fact that we appear to be far from a rigorous understanding of
universality in spectral statistics motivates the development of
heuristic approaches to justifying the random-matrix conjecture in
the expectation that this will lead to further insights into the
reasons underlying its success, and its limitations.  This is our
main purpose here.


A natural starting point for understanding {\it quantum}
consequences of {\it classical} chaos is provided by {\it
semiclassical} techniques. (Other approaches are discussed in Haake
(2001), Muzykantskii \& Khmel'nitskii (1995) and Andreev {\it et
al.} (1996).) By stationary-phase approximation of Feynman's path
integral, Gutzwiller (1990) was able to represent the level density
as a sum over classical periodic orbits. The correlation function
$R(\epsilon)$ is thus a sum over pairs of periodic orbits $a,b$
(Berry 1985). Importantly, this classical double sum incorporates
quantum interference effects through a phase factor ${\rm
e}^{i(S_a-S_b)/\hbar}$ depending on the difference between the two
actions $S_a$ and $S_b$. In the semiclassical limit, as $\hbar\to
0$, constructive interference requires small action differences
$S_a-S_b$, at most of the order of $\hbar$. Constructively
interfering orbit pairs were identified by Hannay \& Ozorio de
Almeida (1984) and Berry (1985), Sieber \& Richter (2001), Sieber
(2002) and M\"uller {\it et al.} (2004, 2005), and shown to yield
non-oscillatory contributions to $R(\epsilon)$: Hannay \& Ozorio de
Almeida (1984) and Berry (1985) showed, using a sum rule that
follows from ergodicity, that `diagonal' pairs of identical or
mutually time-reversed orbits give a contribution quadratic in
$1/\epsilon$ and thus determine $c_2$; Sieber and Richter (2001)
derived the cubic term $c_3$ using pairs of orbits that differ from
each other only by their connections in an `encounter' of two close
orbit stretches; and, building on that insight, it was shown by
M\"uller {\it et al.} (2004, 2005) that higher-order contributions
to $R(\epsilon)$ are due to an infinity of further families of pairs
of orbits differing in arbitrarily many encounters of arbitrarily
many close stretches. Summation of all contributions yields the full
non-oscillatory power series ${\rm Re}\sum_{n=2}^\infty
c_n\left(\frac{1}{\epsilon}\right)^n$. In these calculations the
essential simplifying assumption is that the orbit pairs identified
are the only ones that ultimately contribute.  We shall make the
same assumption in our approach.

The extension to the oscillatory terms proved to be much more
difficult; these terms are related to more subtle correlations
between periodic orbit actions (Argaman {\it et al.} 1993). However, the leading
oscillatory contribution was derived in a modified semiclassical
setting by Bogomoly \& Keating (1996), and an expansion to all orders realized
by Heusler {\it et al.} (2007). The approach of Heusler {\it et al.} 
was based on the following detour inspired by RMT (or more precisely
a field theoretic implementation of RMT, the so-called nonlinear
sigma model): the correlation function $R(\epsilon)$ can be
represented through derivatives of a generating function involving a
ratio of four spectral determinants $\Delta(E)=\det(E-H)$ with four
different energy arguments. (Here $H$ denotes the Hamiltonian.) In
fact, such a representation can be realized in two equivalent ways.
It turned out that with a Gutzwiller-type approximation for the
generating function, one either recovers the non-oscillatory or the
oscillatory terms, depending on which of the two representations is
chosen. This is somewhat paradoxical, because both representations
should in principle give the same result. In each representation a
part of the result is missed because in Gutzwiller's trace formula
the energies have to be taken with sufficiently large imaginary
parts to guarantee convergence, and these imaginary parts can, for
example, cause oscillatory factors to become exponentially small.
Nevertheless, it was observed 
that adding both results one recovers the full
correlation function predicted by RMT. This obviously leaves the
question as to why the two types of contributions are additive, and
why either one by itself does not give the complete answer.

We here justify additivity within the framework of an improved
semiclassical approximation developed by Berry \& Keating (1990),
Keating (1992) and Berry \& Keating (1992).  Their resummation of
the Gutzwiller-type contributions preserves the unitarity of the
time evolution (in particular, the fact that energy eigenvalues are
real) and in doing so reveals a remarkable relationship between
contributions from long and short orbits. It is this relationship
that allows non-oscillatory and oscillatory contributions to be
described naturally within a unified approach.

We shall first review the resummation technique used and illustrate
its application to spectral statistics for an averaged product of
two spectral determinants. We will then turn to the generating
function and the correlation function $R(\epsilon)$, and finally
justify the additivity assumed in Heusler {\it et al.} (2007). To
keep the presentation simple, we will focus on systems without
time-reversal invariance, however the ideas generalize
straightforwardly. Importantly, the treatment of the leading-order
contributions will be carried beyond the universal regime, and so
describes the approach to the semiclassical limit.

\section{Resummation}

To derive a semiclassical approximation for the
spectral determinant one may start from Gutzwiller's formula for
${\rm tr}(E-H)^{-1}$. The spectral determinant is then obtained as
an exponential involving the smoothed number $\overline{N}(E)$ of
energy levels below $E$ and a sum over classical periodic orbits
$a$,
\begin{eqnarray}
\label{gutzwiller_initial} \Delta(E^+)&\propto&\exp
\Big(\int^{E^+}dE'{\rm tr}\frac{1}{E'-H}\Big)\nonumber\\
&\propto&\exp\Big(-i\pi\overline{N}(E^+)-\sum_a F_a {\rm
e}^{iS_a(E^+)/\hbar}\Big).
\end{eqnarray}
Here, $E^+$ denotes an energy with a small positive imaginary part
(needed to ensure convergence), and $S_a$ is the classical action
of the orbit $a$. The factor $F_a$ incorporates an amplitude
depending on the stability of $a$ and the Maslov phase. It is
convenient to expand the exponentiated sum over orbits into a sum
over unordered collections of periodic orbits, or {\it
pseudo-orbits} $A$ (Berry \& Keating 1990)
\begin{eqnarray}
\label{gutzwiller} \Delta(E^+)&\propto&{\rm
e}^{-i\pi\overline{N}(E^+)}\sum_AF_A(-1)^{n_A}{\rm
e}^{iS_A(E^+)/\hbar}\;,
\end{eqnarray}
where $n_A$ is the number of orbits inside $A$ and $S_A$ is the
sum of their actions. $F_A$ is the product of the factors $F_a$ of
the individual orbits (it also includes corrections to the simple
sign factor $(-1)^{n_A}$ that appear if $A$ contains several
identical copies of a given orbit). For negative imaginary parts,
complex conjugation yields
\begin{equation}
\label{gutzwiller_minus} \Delta(E^-)\propto{\rm
e}^{i\pi\overline{N}(E^-)}\sum_AF_A^*(-1)^{n_A}{\rm
e}^{-iS_A(E^-)/\hbar}\;.
\end{equation}

Importantly (\ref{gutzwiller}) and (\ref{gutzwiller_minus}) do not
manifestly preserve the unitarity of the quantum dynamics: in the
limit of vanishing imaginary parts $\Delta(E^+)$ and $\Delta(E^-)$
should become real (given the reality of the energy levels) and
identical to each other. This is not at all obvious from the above
formulas. It was shown in Berry \& Keating (1990), Keating (1992)
and Berry \& Keating (1992) that by explicitly incorporating
unitarity one arrives at an improved approximation for $\Delta(E)$.
Postulating that (\ref{gutzwiller}) and (\ref{gutzwiller_minus})
become identical for $E^+,E^-\to E$ implies duality relations
between pseudo-orbits whose duration (i.e., sum of periods of
contributing orbits) is larger than half of the Heisenberg time
$T_H=2\pi\hbar\overline\rho$ and those shorter than $T_H/2$; the
overall contribution of long orbits is the complex conjugate of the
contribution arising from short orbits. (In the case of billiards
the relationship follows directly from the fact that the spectral
determinant is an even function of the momentum and the spectral
counting function an odd function, see Keating \& Sieber (1994).)
This leads to the improved semiclassical approximation for
$\Delta(E)$ as a sum over pseudo-orbits with durations shorter than
$T_H/2$,
\begin{equation}
\label{resum} \Delta(E)\propto{\rm
e}^{-i\pi\overline{N}(E)}\!\!\sum_{A\; (T_A<T_H/2)}\!\!F_A
(-1)^{n_A}{\rm e}^{i S_A/\hbar} + {\rm c.c.}
\end{equation}
Equation (\ref{resum}) incorporates explicitly the relations between
long and short orbits. Taken together with the orbit correlations
discussed above this leads to a complete semiclassical picture of
spectral fluctuations. It is known as the {\it Riemann-Siegel
lookalike formula} because of its similarity to a corresponding
expression for the Riemann zeta function (Berry \& Keating 1990).

We emphasize that resummation solves an important problem inherent
in (\ref{gutzwiller}) and (\ref{gutzwiller_minus}): whereas
(\ref{gutzwiller}) and (\ref{gutzwiller_minus}) diverge unless the
energy has a sufficiently large imaginary part, the sum in
(\ref{resum}) is finite and so converges when $E$ is real.
Essentially, it is the need to include an imaginary part in the
energy in (\ref{gutzwiller}) and (\ref{gutzwiller_minus}) that is
responsible for the analysis of Heusler {\it et al.} (2007) missing
a part of the RMT expression for the spectral statistics, depending
on the representation used. With (\ref{resum}) we are now able to
access the complete expression.

\section{Product of spectral determinants} To illustrate how
resummation can be used in the context of spectral statistics, we
first consider the product of spectral determinants
\begin{equation}
\label{prod} \Pi(\alpha)=\left\langle\Delta\left(E+\alpha\right)
\Delta\left(E-\alpha\right)\right\rangle\;.
\end{equation}
We shall see that for individual chaotic systems $\Pi(\alpha)$
conforms in the semiclassical limit to the random-matrix prediction
(Ketteman {\it et al.} 1997)
\begin{equation}
\label{prod_RMT}
\Pi(\alpha)=\Pi(0)\sin(2\pi\overline{\rho}\alpha)/2\pi\overline{\rho}\alpha\;.
\end{equation}
To show this, we insert (\ref{resum}) into (\ref{prod}).
$\Pi(\alpha)$ then turns into the following double sum over
pseudo-orbits $A,B$
\begin{eqnarray}
\label{auto_sc}
\Pi(\alpha)&\propto&\Big\langle{\rm
e}^{-i\pi\overline{N}\left(E+\alpha\right)}
\sum_{A\;(T_A<T_H/2)}F_A(-1)^{n_A}{\rm
e}^{iS_A\left(E+\alpha\right)/\hbar}
\nonumber\\
&&\;\;\times{\rm e}^{i\pi\overline{N}\left(E-\alpha\right)}
\sum_{B\;(T_B<T_H/2)}\!\!\!\!\!\!F_B^*(-1)^{n_B}{\rm
e}^{-iS_B\left(E-\alpha\right)/\hbar}\Big\rangle+{\rm c.c.} \;.
\end{eqnarray}
In $\langle\ldots\rangle$ the non-conjugate terms in the
approximation (\ref{resum}) for $\Delta(E+\alpha)$ are paired with
the complex conjugate ones for $\Delta(E-\alpha)$. The term
``+c.c.'' is due to the opposite combination. All other pairings
involve highly oscillatory factors with phases $\propto
\overline{N}(E)$ and thus vanish after averaging over $E$.

Still, the remaining phase factor involving
$\left(S_A\left(E+\alpha\right)-S_B\left(E-\alpha\right)\right)/\hbar$
 oscillates rapidly in the limit $\hbar\to
0$ and thus averages to zero for most pseudo-orbits. Systematic
contributions to (\ref{auto_sc}) therefore arise only from pairs
with action differences at most of the order of $\hbar$ \footnote{
  We neglect contributions where the $S$- and $\overline{N}$-dependent phase
  factors both oscillate rapidly but these oscillations mutually compensate, since
  there is no systematic mechanism giving rise to the pairs of pseudo-orbits needed.}.
In the spirit of the diagonal approximation, the dominating
contributions can be expected to originate from identical
pseudo-orbits $A=B$. If we restrict ourselves to such pairs, and
expand
$\overline{N}\left(E\pm\alpha\right)\approx\overline{N}(E)\pm\overline{\rho}\alpha$,
$S\left(E\pm\alpha\right)\approx S(E)\pm T(E)\alpha$, we obtain
\begin{equation}
\Pi(\alpha)\propto {\rm
e}^{-2i\pi\overline{\rho}\alpha}\!\!\sum_{A\;(T_A<T_H/2)}\!\!
|F_A|^2{\rm e}^{2i\alpha T_A/\hbar}+{\rm c.c.}
\end{equation}

It is now tempting to drop the upper limit $T_H/2$ since it tends to
$\infty$ in the semiclassical limit and, given the factor $|F_A|^2$,
it is no longer needed for convergence (for $\alpha$ with an {\it
arbitrarily small} imaginary part). Indeed, dropping the upper bound
is justified immediately if semiclassically
$\alpha\overline{\rho}\rightarrow\infty$. The result can then be
written as
\begin{equation}
\label{auto_zeta} \Pi(\alpha)\propto{\rm
e}^{-2i\pi\overline{\rho}\alpha}\zeta^{-1}(-2i\alpha/\hbar)+{\rm
c.c.}
\end{equation}
where the dynamical zeta function $\zeta(s)$ is defined by
\begin{equation}
\zeta(s)=\sum_A |F_A|^2(-1)^{n_A}{\rm e}^{-sT_A}
\end{equation}
or, equivalently,
\begin{equation}
\zeta^{-1}(s)=\sum_A |F_A|^2{\rm e}^{-sT_A}.
\end{equation}
Importantly, in chaotic systems $\zeta(s)$ has a simple zero at
$s=0$ (Haake 2001).  This is equivalent to the periodic orbit sum
rule that follows from classical ergodicity (Hannay \& Ozorio de
Almeida 1984).

For energy differences of the order of the mean level spacing the
same result is obtained after a short calculation: if we
incorporate the condition $T_A<T_H/2$ through a step-function
$\Theta\left(x\right)=\frac{1}{2\pi
i}\int_{c-i\infty}^{c+i\infty}\frac{dz}{z} {\rm e}^{xz/\hbar}$
(with small positive $c$) $\Pi(\alpha)$ can be written as
\begin{eqnarray}
\Pi(\alpha)&\propto&{\rm
e}^{-2i\pi\overline{\rho}\alpha}\frac{1}{2\pi
i}\int_{c-i\infty}^{c+i\infty}
\frac{dz}{z}{\rm e}^{\pi\overline{\rho}z}\sum_{A}|F_A|^2{\rm e}^{\frac{2i\alpha-z}{\hbar}T_A}+{\rm c.c.}\nonumber\\
&=&{\rm
e}^{-2i\pi\overline{\rho}\alpha}\frac{1}{2\pi
i}\int_{c-i\infty}^{c+i\infty} \frac{dz}{z}{\rm
e}^{\pi\overline{\rho}z}
\zeta\left(\frac{z-2i\alpha}{\hbar}\right)^{-1}+{\rm c.c.}
\end{eqnarray}
If we close the contour on the left, the residue at $z=0$ reproduces
(\ref{auto_zeta}) and the residue due to the zero of $\zeta(s)$ at
$s=0$ is proportional to $(2i\alpha)^{-1}+{\rm c.c.}=0$ and thus
vanishes.

We note that equation (\ref{auto_zeta}) includes non-universal
corrections to the random-matrix prediction (\ref{prod_RMT}),
encoded in the analytic structure of $\zeta(s)$. It reduces to
(\ref{prod_RMT}) in the semiclassical limit because
$\zeta(s)\propto s$ as $s\rightarrow 0$; that is, as a consequence
of classical ergodicity. We have thus seen how oscillatory
contributions to spectral statistics arise naturally from a
resummed semiclassical approximation, even beyond the universal
regime.

\section{Generating function}

Let us now consider the generating
function for $R(\epsilon)$,
\begin{eqnarray}
Z(\alpha,\beta,\gamma,\delta)\equiv\left\langle\frac{\Delta(E+\gamma)\Delta(E-\delta)}{\Delta(E+\alpha)\Delta(E-\beta)}\right\rangle\;.
\end{eqnarray}
Here the two energy differences $\alpha,\beta$ in the denominator
are taken with small positive imaginary parts. Then
\begin{eqnarray}
\label{deriv} R(\epsilon)=-\frac{1}{2\pi^2\overline{\rho}^2}{\rm
Re}\frac{\partial^2
Z}{\partial\alpha\partial\beta}\big|_{(\|)}-\frac{1}{2}
\end{eqnarray}
where $(\|)$ indicates the limit
$\alpha,\beta,\gamma,\delta\to\frac{\epsilon}{2\pi\overline\rho}$.
Equation (\ref{deriv}) is easily checked: taking two derivatives
of $Z$ leads to ${\rm tr}\frac{1}{E+\alpha-H}{\rm
tr}\frac{1}{E-\beta-H}$ times the original ratio of four
determinants; the two traces ultimately lead to level densities
whereas the ratio converges to unity in the limit $(\|)$.

To obtain a resummed semiclassical approximation for $Z$ we use the
Riemann-Siegel lookalike (\ref{resum}) for the two determinants in
the numerator. Due to the important role played by the imaginary
parts in the denominator, no resummation is possible there, and we
rather stick to the unresummed expressions
\begin{equation}
\Delta(E^+)^{-1}\propto{\rm e}^{i\pi\overline{N}(E^+)}\sum_AF_A{\rm
e}^{iS_A(E^+)/\hbar}\;
\end{equation}
and
\begin{equation}
\Delta(E^-)^{-1}\propto{\rm
e}^{-i\pi\overline{N}(E^-)}\sum_AF_{A}^*{\rm e}^{-iS_A(E^-)/\hbar}.
\end{equation}
Specifically, no resummation of $\Delta^{-1}$ as a finite sum can
reproduce the poles at the energy levels (a finite sum can, of
course, reproduce the corresponding zeros of $\Delta$ itself). The
choice of when we use $\Delta(E^+)$ or $\Delta(E^-)$ is fixed by the
positions of these poles. Collecting the pseudo-orbit sums from all
four spectral determinants we then obtain two sums over quadruplets
of pseudo-orbits $A,B,C,D$,
\begin{eqnarray}
\label{Z_zeta} &&Z=\left\langle{\rm
e}^{i\pi\overline{N}(E+\alpha)}\sum_AF_A{\rm
e}^{iS_A(E+\alpha)/\hbar}\right.
\nonumber\\
&&\times{\rm e}^{-i\pi\overline{N}(E-\beta)}\sum_BF_B^*{\rm
e}^{-iS_B(E-\beta)/\hbar}
\nonumber\\
&&\times{\rm e}^{-i\pi\overline{N}\left(E+\gamma\right)}
\!\!\sum_{C\;(T_C<T_H/2)}\!\!F_C(-1)^{n_C}{\rm
e}^{iS_C\left(E+\gamma\right)/\hbar}
\nonumber\\
&&\left.\times{\rm e}^{i\pi\overline{N}\left(E-\delta\right)}
\!\!\sum_{D\;(T_D<T_H/2)}\!\!F_D^*(-1)^{n_D}{\rm
e}^{-iS_D\left(E-\delta\right)/\hbar}\right\rangle\nonumber\\
&&+\{ \gamma\to-\delta,\;\delta\to-\gamma\} \;.
\end{eqnarray}
These sums arise from combining the non-conjugate terms in
(\ref{resum}) for $\Delta(E+\gamma)$ with the complex conjugate
ones for $\Delta(E-\delta)$ and vice versa.

Interference between pseudo-orbits leads to a phase factor
\begin{equation}
{\rm
e}^{i(S_A(E+\alpha)-S_B(E-\beta)+S_C(E+\gamma)-S_D(E-\delta))/\hbar}\;.
\end{equation}
The pseudo-orbits will interfere constructively if the cumulative
action of $A$ and $C$ nearly coincides with the cumulative action of
$B$ and $D$. The simplest way to realize this is to have the orbits
in $B$ and $D$ identical to those in $A$ and $C$ (neglecting
repetitions). The sum over such ``diagonal'' quadruplets of
pseudo-orbits can be split into sums over the intersections $A\cap
B,A\cap D,C\cap B,C\cap D$. If we expand $S$ and $\overline{N}$ as
above these sums can be written as
\begin{eqnarray}
Z&=&{\rm
e}^{i\pi\overline{\rho}(\alpha+\beta-\gamma-\delta)}\!\!\!\!\!\!\!\!\!\!\!\sum_{A\cap
B,A\cap D,C\cap B,C\cap
D\,\atop(T_C,T_D<T_H/2)}\!\!\!\!\!\!\!\!\!\!\!
|F_{A\cap B}|^2{\rm e}^{iT_{A\cap B}(\alpha+\beta)/\hbar}\nonumber\\
&\times&|F_{A\cap D}|^2(-1)^{n_{A\cap D}}{\rm e}^{iT_{A\cap D}(\alpha+\delta)/\hbar}\nonumber\\
&\times&|F_{C\cap B}|^2(-1)^{n_{C\cap B}}{\rm e}^{iT_{C\cap B}(\gamma+\beta)/\hbar}\nonumber\\
&\times&|F_{C\cap D}|^2{\rm e}^{iT_{C\cap
D}(\gamma+\delta)/\hbar}+\{
\gamma\!\to\!-\delta,\;\delta\!\to\!-\gamma\}.
\end{eqnarray}
In the same way as for the product of spectral determinants, we now
drop the upper limits at $T_H/2$\footnote{Again, convergence is
guaranteed when $\alpha, \beta, \gamma$ and $\delta$ are given {\it
arbitrarily small} imaginary parts.}. This is clearly justified if
semiclassically $\alpha\overline{\rho}\rightarrow\infty$ etc., and
there are no singularities prohibiting analytic continuation to
$\alpha,\beta,\gamma,\delta\propto 1/\overline{\rho}$. Evaluating
the sum in that case gives
\begin{eqnarray}
\label{final}
 Z&=&{\rm
e}^{i\pi\overline{\rho}(\alpha+\beta-\gamma-\delta)}\frac{\zeta(-i(\alpha+\delta)/\hbar)\zeta(-i(\gamma+\beta)/\hbar)}
{\zeta(-i(\alpha+\beta)/\hbar)\zeta(-i(\gamma+\delta)/\hbar)}+\{ \gamma\to-\delta,\;\delta\to-\gamma\}\;.
\end{eqnarray}

Like (\ref{auto_zeta}), equation (\ref{final}) is valid even
beyond the universal regime; that is, it describes the
system-specific approach to the semiclassical limit via the
analytic structure of the classical zeta function $\zeta(s)$.
Taking the semiclassical limit we find
\begin{eqnarray}
Z&=&{\rm
e}^{i\pi\overline{\rho}(\alpha+\beta-\gamma-\delta)}\frac{(\alpha+\delta)(\gamma+\beta)}
{(\alpha+\beta)(\gamma+\delta)}
+\{ \gamma\to-\delta,\;\delta\to-\gamma\}\equiv Z_1+Z_2\nonumber\\
R(\epsilon)&=&-\frac{1}{2\pi^2\overline{\rho}^2}{\rm
Re}\frac{\partial^2
Z}{\partial\alpha\partial\beta}\big|_{(\|)}-\frac{1}{2}
=-\left(\frac{\sin\epsilon}{\epsilon}\right)^2
\end{eqnarray}
which is exactly the random-matrix prediction for the GUE
($c_2=-1/2, d_2=1/2$). Here the summands $Z_1$ and $Z_2$ are
responsible for the non-oscillatory and oscillatory contributions to
the correlation function, corresponding to two different saddle
points in RMT. We stress that $Z_2$ becomes accessible to a
semiclassical treatment only through resummation.  Differentiating
(\ref{final}) gives a formula identical to that due to Bogomolny \&
Keating (1996) for the nonuniversal corrections to the random-matrix
expression for $R(\epsilon)$.

\section{Relation to Heusler {\it et al.}} 
In Heusler {\it et al.} (2007) no resummation was made and thus only
the first summand $Z_1$ could be obtained. Hence if one represents
$R(\epsilon)$ as in (\ref{deriv}) only non-oscillatory contributions
are found. To recover oscillatory contributions an alternative
representation of $R(\epsilon)$ through $Z$ was used, with the limit
$(\|)$ substituted by $(\times)$
$\alpha,\beta\to\frac{\epsilon}{2\pi\overline\rho}$,
$\gamma,\delta\to-\frac{\epsilon}{2\pi\overline\rho}$. If one
replaces $Z$ by $Z_1$ (and drops $-\frac{1}{2}$) this time only
oscillatory contributions are obtained. It was claimed that
summation of both results
\begin{equation}
-\frac{1}{2\pi^2\overline{\rho}^2}\frac{\partial^2
Z_1}{\partial\alpha\partial\beta}\big|_{(\|)}
-\frac{1}{2\pi^2\overline{\rho}^2}\frac{\partial^2
Z_1}{\partial\alpha\partial\beta}\big|_{(\times)} -\frac{1}{2}
\end{equation}
yields the full correlation function. After comparison to
(\ref{deriv}) we see that this claim is equivalent to
\begin{equation}
\label{parcross} \frac{\partial^2
Z_1}{\partial\alpha\partial\beta}\big|_{(\times)}= \frac{\partial^2
Z_2}{\partial\alpha\partial\beta}\big|_{(\|)}
\end{equation}
which is easily verified from Riemann-Siegel resummation: the
right-hand side of (\ref{parcross}) differs from $\frac{\partial^2
Z_1}{\partial\alpha\partial\beta}\big|_{(\|)}$ because $\gamma$
and $\delta$ are interchanged and flipped in sign; on the left the
same effect is reached by replacing $(\|)$ with $(\times)$.

\section{Conclusions and outlook}

By resummation of the generating
function we have clarified the semiclassical origin of oscillatory
contributions to spectral correlators. This opens the door for a
systematic use of generating functions in semiclassical theory, with
possible applications in quantum transport or to higher-order
correlation functions. Moreover we have extended the semiclassical
treatment beyond the universal regime. Within
the diagonal approximation, non-universal effects come into play
through the dynamical zeta function. It would be interesting to see
if this carries over to off-diagonal terms.

\begin{acknowledgements}

We are grateful to A. Altland, P. Braun, F. Haake, S. Heusler, M. Sieber and B.
Simons for helpful discussions.

\end{acknowledgements}

\label{lastpage}

\end{document}